%% file: main.tex
\documentclass[11pt,oneside,a4paper]{article}
\usepackage{xparse}
\usepackage{amsmath}
\usepackage{amssymb}
\usepackage[english]{babel}
\usepackage[utf8]{inputenc}
\usepackage[]{nicefrac}
\usepackage{amsmath, amssymb, amsthm}
\usepackage{graphicx}
\usepackage[colorinlistoftodos]{todonotes}
\usepackage{tikz}
\usepackage[affil-it]{authblk}
\usepackage[]{ccicons}
\usepackage[]{xcolor}
\usepackage{leftidx}
\usepackage{natbib}
\usepackage{pgfplots}
\usepackage[]{hyperref}
\hypersetup{%
    colorlinks = true,
    allcolors = gray,
}

\newtheorem{remark}{Remark}[section]

\newcommand{\dVblank}{\, {\rm d}V}

\NewDocumentCommand \dV{ o }{
    \IfNoValueTF{#1}{\dVblank}
    {
        \dV_{#1}
    }
}

\NewDocumentCommand \eref{ m }{
    (\ref{eqn:#1})
}

\NewDocumentCommand \fref{ m }{
    fig.~\ref{fig:#1}
}

\NewDocumentCommand \an{ m }{
    \langle {#1} \rangle
}

\NewDocumentCommand \mbf{ m }{
    \mathbf{#1}
}

\NewDocumentCommand \stateOne{ m }{
    \underline{\mbf{#1}}
}

\NewDocumentCommand \stateTwo{ m o }{
    \IfNoValueTF{#2}{\stateOne{#1}}
    {
    \stateOne{#1}\an{#2}
    }
}

\NewDocumentCommand \s{ m o o }{
    \IfNoValueTF{#3}{\stateTwo{#1}[#2]}
    {
    \underline{\mbf{#1}}(\mbf{#3})\an{#2}
    }
}

\NewDocumentCommand \stateOneI{ m m }{
    \underline{#1}_{#2}
}

\NewDocumentCommand \stateTwoI{ m m o }{
    \IfNoValueTF{#3}{\stateOneI{#1}{#2}}
    {
    \stateOneI{#1}{#2}\an{#3}
    }
}

\NewDocumentCommand \si{ m m o o }{
    \IfNoValueTF{#4}{\stateTwoI{#1}{#2}[#3]}
    {
        \stateOneI{#1}{#2}(\mbf{#4})\an{#3}
    }
}

\title{Deriving peridynamic influence functions for one-dimensional elastic materials with periodic microstructure.}
\date{\today}
\author{Xiao Xu\thanks{xiaoxu42@utexas.edu}
\and John T. Foster\thanks{john.foster@utexas.edu} \\ 
The Oden Institute for Computational Engineering and Sciences \\ 
The University of Texas at Austin}

\begin{document}
\maketitle
\begin{abstract}
    The influence function in peridynamic material models has a large effect on the dynamic behavior of elastic waves and in turn can greatly effect dynamic simulations of fracture propagation and material failure. Typically, the influence functions that are used in peridynamic models are selected for their numerical properties without regard to physical considerations. In this work, we present a method of deriving the peridynamic influence function for a one-dimensional initial/boundary-value problem in a material with periodic microstructure. Starting with the linear local elastodynamic equation of motion in the microscale, we first use polynomial anzatzes to approximate microstructural displacements and then derive the homogenized nonlocal dynamic equation of motion for the macroscopic displacements; which, is easily reformulated as linear peridyamic equation with a discrete influence function. The shape and localization of the discrete influence function is completely determined by microstructural mechanical properties and length scales. By comparison with a highly resolved microstructural finite element model and the standard linear peridynamic model with a linearly decaying influence function, we demonstrate that the influence function derived from microstructural considerations is more accurate in predicting time dependent displacements and wave dynamics.
\end{abstract}
\input{Introduction}
\input{Derivation}
\input{Numerical}
\input{Conclusion}

\bibliographystyle{abbrvnat}
\bibliography{mybib}
\end{document}

%% file: Introduction.tex
\section{Introduction}
Peridynamics was first proposed as a reformulation of the classical continuum linear momentum balance law by \citet{silling2000reformulation}. This nonlocal model replaces the spatial derivatives in the classical conservation of momentum equation with an integral functional to determine the net internal force density on a material point. The integral formulation has advantages over the classical theory when solving problems with discontinuities like cracks and material fragmentation. A generalization of the original formulation, often called state-based peridynamics \cite{silling2007psa}, introduced the concept of an \emph{influence function} into the constitutive model.  The influence function weights the individual interactions of each pair of material points in a peridynamic body.  The influence function has similarities with the kernel functions, window functions, or weight functions in convolution, and interpolation techniques on scattered data.  When referring to the part of the integrand that weights contributions from the variable of integration without contribution from material parameters we will use the terminology \emph{influence function}, and when referring to the kernel of an integral in the standard sense of convolution, we will use the terminology \emph{kernel function}. While peridynamics has been used to model complex material behavior \cite{warren2009non, silling2010peridynamic, foster2010viscoplasticity}, and has shown unique capabilities in numerical simulation of crack propagation \cite{agwai2011predicting}, crack branching \cite{ha2010studies, bobaru2015cracks} and damage in composite laminates \cite{xu2008peridynamic,lai2015peridynamics}, what has been missing in the peridynamic literature is a systematic way to determine the peridynamic influence function. This is despite the knowledge that it is  key factor contributing to the behavior of peridynamic material models \cite{weckner2005effect}, especially in the presence of fracture \cite{seleson2011role}.  A common question raised by those curious about peridynamics is, ``How do you chose the peridynamic horizon?''.  As we show in this paper, a more appropriate question should be, ``How does one construct the peridynamic kernel function?\footnote{As the kernel function can be localized at any length scale, the first question is embedded in the second.}''. 

For nonlocal flow in porous media, \citet{delgoshaie2015non} used the multiscale connectivity of natural pore networks to explain anomalous diffusive behavior and used pore network mesoscale computational models to extract nonlocal kernal functions for use in continuum models. With respect to continuum solid mechanics, there is not always the physical existence of long-range forces between material points which makes arguments for nonlocality more challenging. \citet{silling2014origin} demonstrated that nonlocality in solid materials can arise from the small-scale heterogeneity that is excluded through a implicit or explicit homogenization procedure. This suggests that the peridynamic kernel function for solid materials should be related to the microstructure of the solid. Motivated by this idea, we present a theoretical way to compute the discrete peridynamic kernel function for one-dimensional elasticity from a given heterogeneous microstructure.

We focus on a simple one-dimensional initial/boundary-value problem in a material with periodic heterogeneous microstructure. Polynomial ansatzes are used for microstructure displacements, and the peridynamic kernel function is calculated based on the mechanics of the microstructure. Then, the nonlocal elastodynamics for the macroscopic (average) displacement is derived and solved. The resulting formulation is more accurate in resolving wave dynamics for this model problem when compared to standard influence functions used in peridynamic analysis in the literature.

The rest of the paper is organized as follows: {\S\ref{chap:problem}} presents the setting of the elastodynamic problem with periodic heterogeneity and the definitions for the multiscale quantities of interest. {\S\ref{chap:multi}} demonstrates how we bridge macroscopic quantities with microstructure, and the derivations for the discrete peridynamic kernel function. In {\S\ref{chap:higher}} we outline a higher-order approximation for more accurate results. {\S\ref{chap:numerical}} presents numerical simulations using the discrete peridynamic kernel functions and comparisons with the results generated by a standard peridynamic model and highly resolved classical finite element methods. 

%% file: Derivation.tex
\section{One-dimensional elastodynamic composite problem}

\label{chap:problem}
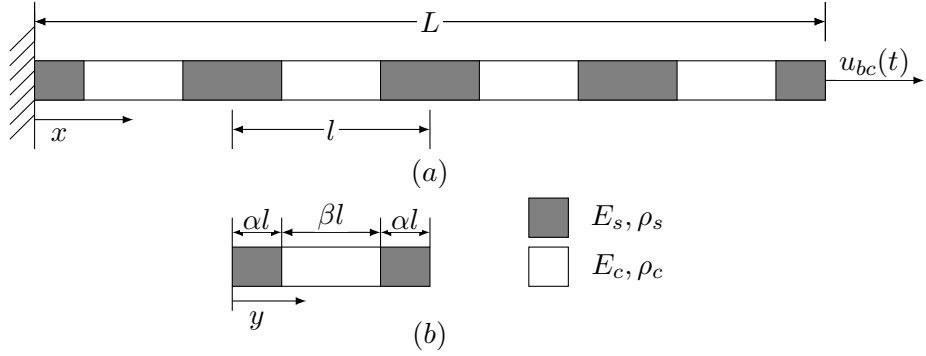
\begin{figure}
\centering
\begin{tikzpicture}[scale=0.65]
    \tikzstyle{pt}=[circle, fill=black, inner sep=0pt, minimum size=5pt]
    \draw (-8,0) -- (8,0) -- (8,0.8) -- (-8,0.8);
    \draw (-8,-1) -- (-8,1.8);
    \draw (-8,1.5) -- (-8.5,1);
    \draw (-8,1.2) -- (-8.5,0.7);
    \draw (-8,0.9) -- (-8.5,0.4);
    \draw (-8,0.6) -- (-8.5,0.1);
    \draw (-8,0.3) -- (-8.5,-0.2);
    \draw (-8,0) -- (-8.5,-0.5);
    \draw (-8,-0.3) -- (-8.5,-0.8);
    
	\draw[-latex] (-8,-0.4) -- (-6,-0.4); 		  
    \draw (-7.5,-0.7) node() {$x$};
    \draw[-latex] (8,0.4) -- (10,0.4) node[midway, above , fill=white, inner sep=1pt] {$u_{bc}(t)$} ;
    
    \draw[latex-latex] (8,1.6) -- (-8,1.6) node[midway , fill=white, inner sep=1pt] {$L$} ;
    \draw (-8,1.2) -- (-8,2);
    \draw (8,1.2) -- (8,2);
    
    \filldraw[fill=black!50, draw=black] (-8,0) rectangle (-7,0.8);
    \filldraw[fill=black!50, draw=black] (-5,0) rectangle (-3,0.8);
    \filldraw[fill=black!50, draw=black] (-1,0) rectangle (1,0.8);
    \filldraw[fill=black!50, draw=black] (3,0) rectangle (5,0.8);
    \filldraw[fill=black!50, draw=black] (7,0) rectangle (8,0.8);
   \draw (-4,-0.2) -- (-4,-1);
   \draw (0,-0.2) -- (0,-1);
   \draw[latex-latex] (-4,-0.6) -- (0,-0.6) node[midway , fill=white, inner sep=1pt] {$l$} ;
   \draw (0,-1.5) node() {$(a)$};
   \filldraw[fill=black!50, draw=black] (-4,-3) rectangle (-3,-3.8);
    \filldraw[fill=black!50, draw=black] (-1,-3) rectangle (0,-3.8);
    \draw (-3,-3.8) -- (-1,-3.8);
    \draw (-3,-3) -- (-1,-3);
    \draw (-4,-3.8) -- (-4,-4.3);
    \draw[-latex] (-4,-4.1) -- (-2.5,-4.1); 		  
    \draw (-3.5,-4.5) node() {$y$};
    
    \draw[latex-latex] (-4,-2.7) -- (-3,-2.7) node[midway , above,  fill=white, inner sep=1pt] {$\alpha l$} ;
    \draw (-4,-3) -- (-4,-2.4);
    \draw (-3,-3) -- (-3,-2.4);
    
    \draw[latex-latex] (0,-2.7) -- (-1,-2.7) node[midway , above,  fill=white, inner sep=1pt] {$\alpha l$} ;
    \draw (0,-3) -- (0,-2.4);
    \draw (-1,-3) -- (-1,-2.4);
    
    \draw[latex-latex] (-3,-2.7) -- (-1,-2.7) node[midway , above,  fill=white, inner sep=1pt] {$\beta l$} ;
    
    \filldraw[fill=black!50, draw=black] (2,-2) rectangle (2.8,-2.8);
    \draw (4,-2.4) node() {$ E_s,\rho_s$};

    \filldraw[fill=white, draw=black] (2,-3) rectangle (2.8,-3.8);
    \draw (4,-3.4) node() {$ E_c,\rho_c$};

    \draw[latex-latex] (8,1.6) -- (-8,1.6) node[midway , fill=white, inner sep=1pt] {$L$} ;
    \draw (-8,1.2) -- (-8,2);
    \draw (8,1.2) -- (8,2);
    
    \draw (0,-4.8) node() {$(b)$};

\end{tikzpicture}
\caption{$(a)$ One-dimensional periodic composite  $(b)$ Unit cell with heterogeneity}
\label{fig:Illustration}
\end{figure}

Here, we consider the one-dimensional elastic composite problem inspired by \citep{fish2001higher}, a composite rod composed of a periodic array of two linearly elastic, homogeneous, and isotropic constituents with perfect interfaces as shown in Figure~\ref{fig:Illustration}. The composite is fixed at one end and subjected to an axial time-dependent displacement boundary condition $u_{bc}(t)$ at the other. It has a macroscopic coordinate $x$ that originates at the fixed end. The total length of the composite rod is $L$ and each point at macroscopic coordinates $x$ has a microscopic unit cell where $y$ denotes its microscopic coordinate. The dark block represents the stiffer constituent with elastic modulus $E_s$ and density $\rho_s$ while the white block represents the more compliant constituent with elastic modulus $E_c$ and density $\rho_c$. In order to keep the symmetry of the microscopic unit cell, we define the compliant constituent be in the middle of the unit cell with length $\beta l$, while the stiff constituents are at each end of the unit cell with length $\alpha l$. The consistency of microscopic geometry requires
\[2\alpha + \beta = 1. \]
The displacement of every point in the microstructure of the composite can be expressed using both its macroscopic coordinate $x$ and microscopic coordinate $y$ as $u(x,y)$. It is reasonable to define the macroscopic displacement $u(x)$ as the average microstructural displacement inside the unit cell at the macroscopic coordinate $x$
\begin{align}
	u(x)  = \frac 1 l \int^l_0 u(x,y) {\rm d} y. \label{eqn:averagedisplacement}
\end{align}
Conservation of linear momentum at the microscopic scale is 
\begin{subequations}
\begin{align}
	\rho(y) \ddot{u}(x,y,t) = \sigma(x,y,t)_{,y}, \label{eqn:governingequations1}
\end{align}
with
\begin{align}
    \sigma(x,y,t) = E(y)u(x,y,t)_{,y}, \label{eqn:governingequations1}
\end{align}
\label{eqn:governingequations}
\end{subequations}
where $u(x,y,t)$ and $\sigma(x,y,t)$ are the axial displacement and stress at the microscopic scale, respectively. The double dot above $u$ indicates two time derivatives and the subscripts following the comma indicate spatial differentiation with respect to the subscripted variable following standard mathematical notation conventions.  $\rho(y)$ and $E(y)$ are the density and elastic modulus in the microscopic unit cell, which in this case are
\begin{align}
    \rho(y) &= 
    \begin{cases}
        \rho_s & 0<y<\alpha l, \quad  (\alpha + \beta)l<y<l,\\
        \rho_c & \alpha l \le y \le (\alpha +\beta )l, 
    \end{cases} \notag \\
    E(y) &= 
    \begin{cases}
        E_s & 0<y<\alpha l, \quad (\alpha + \beta)l<y<l,\\
        E_c & \alpha l \le y \le (\alpha +\beta )l,
    \end{cases} \notag
\end{align}
We combine \eref{averagedisplacement} and \eref{governingequations} to write the acceleration of macroscopic displacement $\ddot{u}(x,t)$
\begin{align}
    \ddot{u}(x,t) &= \frac 1 l \int^l_0 \ddot{u}(x,y,t) {\rm d} y, \notag \\
    &= \frac 1 l \int^l_0 \frac{(E(y)u(x,y,t)_{,y})_{,y}}{\rho(y)} {\rm d}y.
    \label{eqn:TDaveragedisplacement}
\end{align}

\section{Multiscale analysis}
\label{chap:multi}
As shown in \fref{Illustration}, there are three interfaces inside a unit cell given its periodicity. Therefore, the continuity of displacement and stress across these three interface requires the following conditions
\begin{subequations}
\begin{align}
    u(x,\alpha l^-) &= u(x, \alpha l^+ )  \\
    E_s u(x,\alpha l^- )_{,y} &= E_c u(x,\alpha l^+)_{,y}  \\
    u(x,\alpha l+ \beta l^-) &= u(x, \alpha l +\beta l^+ )  \\
    E_c u(x,\alpha l+\beta l^- )_{,y} &= E_s u(x,\alpha l+\beta l^+)_{,y}  \\
	u(x+l,0,t)  &= u(x,l,t)   \\
    E_s u(x+l,0,t)_{,y} &= E_s u(x,l,t)_{,y} 
\end{align}
\label{eqn:continuity1}
\end{subequations}
where the $+/-$ superscripts represent the right side of the interface and the left side of the interface, respectively.

For the purposes of demonstration, we adopt the following anzatzes for the microstructural displacement, for the stiff constituent inside the unit cell we assume a quadratic displacement field and for the compliant constituent we assume the displacement is cubic. We will generalize the theory to arbitrary order polynomials for more accuracy in the sequel. Therefore, the microstructural displacement at the macroscopic coordinate $x$ according to our assumptions can be written as
\begin{align}
u(x,y) = 
\begin{cases}
    m(x)y^2 + a(x)y +b(x) & 0<y<\alpha l, \\
    r(x)(y-\alpha l)^3 + c(x)(y - \alpha l)^2 & \\
    \quad + d(x)(y-\alpha l) + e(x)  &  \alpha l < y< (\alpha + \beta)l, \\
    n(x)(y-\alpha l-\beta l)^2 & \\ 
    \quad + f(x)(y-\alpha l-\beta l) +g(x) &  (\alpha+\beta)l<y<l,
\end{cases}
\label{eqn:cubicquadratic}
\end{align}
where $a(x)$, $b(x)$, $c(x)$, $d(x)$, $e(x)$, $f(x)$, $g(x)$, $m(x)$, $n(x)$ and $r(x)$ are coefficients that are only dependent on the macroscopic coordinate $x$. Determining these coefficients will allow us to use \eref{TDaveragedisplacement} to define the elastodynamic problem in terms of the macroscopic displacement. 

In order to determine the coefficients, we will require one additional order of continuity at the interfaces. To demonstrate, consider the interface at $y=\alpha l$, we'll require
\begin{align}
    \ddot{u}(x,\alpha l^-) &= \ddot{u}(x,\alpha l^+), \notag
\end{align}
and using \eref{governingequations} we have
\begin{subequations}
\begin{align}
    \frac{E(\alpha l^-)} {\rho(\alpha l^-)} u(x,\alpha l^-)_{,yy} &= \frac{E(\alpha l^+)}{\rho(\alpha l^+)}u(x,\alpha l^+)_{,yy}.
\end{align}
Following this assumption, we then have two more equations of continuity inside a unit cell
\begin{align}
    \frac{E_c} {\rho_c} u(x,\alpha l+\beta l^-)_{,yy}=& \frac{E_s}{\rho_s}u(x,\alpha l+\beta l^+)_{,yy},  \\
    \frac{E_s} {\rho_s} u(x+l,0)_{,yy}=& \frac{E_s}{\rho_s}u(x, l )_{,yy}.
\end{align}
\label{eqn:extracontinuity}
\end{subequations}
Substituting \eref{cubicquadratic} into \eref{averagedisplacement} and integrating along with the interface continuity equations \eref{continuity1} and \eref{extracontinuity} gives
\begin{align}
    l\bar u(x) &= \frac{\alpha^3 l^3(m(x)+n(x))}{3} + \frac{\alpha^2 l^2(a(x)+f(x))}{2} + \alpha l(b(x)+g(x))\notag \\
               & \quad + \frac{c(x)\beta^3 l^3}{3} +  \frac{d(x)\beta^2 l^2}{2} + e(x)\beta l + \frac{r(x)\beta^4 l^4}{4},  \notag \\
b(x+l) &= n(x)\alpha^2 l^2 + f(x)\alpha l  +g(x), \notag \\
a(x+l) &= f(x) + 2n(x)\alpha l, \notag \\
m(x+l) &= n(x), \notag \\
 e(x) &= m(x)\alpha^2l^2 + a(x)\alpha l + b(x), \notag \\
 E_c d(x) &= E_s (2m(x)\alpha l + a(x)), \notag \\
\frac{E_c} {\rho_s}c(x) &= \frac{E_s}{\rho_c}m(x), \notag \\
g(x) &= r(x)\beta^3 l^3 + c(x)\beta^2 l^2 + d(x)\beta l +e(x), \notag \\
E_s f(x) &=  E_c ( 3r(x)\beta^2 l^2 + 2c(x)\beta l + d(x)), \notag \\
\frac{E_s}{\rho_s}n(x) &= \frac{E_c}{\rho_c} (3r(x)\beta l + c(x)) \notag
\end{align}
Fourier transforming each of the equations gives 
\begin{align}
    l U(\xi) &= \frac{\alpha^3 l^3(M(\xi)+N(\xi))}{3} + \frac{\alpha^2 l^2(A(\xi)+F(\xi))}{2} + \alpha l(B(\xi)+G(\xi))  \notag \\
& \quad + \frac{C(\xi)\beta^3 l^3}{3} +  \frac{D(\xi)\beta^2 l^2}{2} + E(x)\beta l + \frac{R(x)\beta^4 l^4}{4}, \notag \\
e^{i2\pi l\xi}B(\xi) &= N(\xi)\alpha^2 l^2 + F(\xi)\alpha l  + G(\xi), \notag \\
e^{i2\pi l\xi}A(\xi) &= F(\xi) + 2N(\xi)\alpha l,\notag \\
e^{i2\pi l\xi}M(\xi) &= N(\xi), \notag \\
 E(\xi) &= M(\xi)\alpha^2l^2 + A(\xi)\alpha l + B(\xi), \notag \\
 E_c D(\xi) &= E_s (2M(\xi)\alpha l + A(\xi)),\notag \\
E_c C(\xi) &= E_s M(\xi); \notag \\
G(\xi) &= R(\xi)\beta^3 l^3 + C(\xi)\beta^2 l^2 + D(\xi)\beta l +E(\xi),\notag \\
E_s F(\xi) &=  E_c ( 3R(\xi)\beta^2 l^2 + 2C(\xi)\beta l + D(\xi)), \notag \\
E_s N(\xi) &= E_c(3R(\xi)\beta l + C(\xi)), \notag
\end{align}
where the uppercase function symbols are used to represent the Fourier transform of the corresponding lowercase symbols. Now solve for $A(\xi)$
\begin{align}
A(\xi) &= \frac{12(e^{i2\pi l\xi}-1)( 2\alpha \frac{\rho_s}{\rho_c}+\beta) U(\xi)}{l^2( a_0  +a_1e^{i2\pi l\xi} +a_2e^{i4\pi l\xi} )   }, \label{eqn:A_sol}
\end{align}
where $a_0,a_1,a_2$ are dimensionless coefficients determined by material properties
\begin{subequations}
\begin{align}
    a_0 &= 4 \frac{E_{h}}{E_c} \alpha \beta^{2}  +  \frac{E_{h}}{E_c} \beta^{3}  + 4 \alpha^{3} \frac{\rho_{h}}{\rho_c} + 6  \alpha^{2} \beta , \\
    a_1 &= 48\frac{E_{h}}{E_c}\alpha^{2} \beta \frac{\rho_{h}}{\rho_c} + 24  \frac{E_{h}}{E_c}\alpha \beta^{2} \frac{\rho_{h}}{\rho_c} +  16 \frac{E_{h}}{E_c} \alpha \beta^{2}   +  10 \frac{E_{h}}{E_c} \beta^{3} \notag \\
    & \quad + 88 \alpha^{3} \frac{\rho_{h}}{\rho_c} + 48  \alpha^{2} \beta \frac{\rho_{h}}{\rho_c} + 36  \alpha^{2} \beta  + 24  \alpha \beta^{2} , \\
    a_2 &=  4 \frac{E_{h}}{E_c} \alpha \beta^{2}  +  \frac{E_{h}}{E_c} \beta^{3}  + 4 \alpha^{3} \frac{\rho_{h}}{\rho_c} + 6  \alpha^{2} \beta = a_0.
\end{align}
\label{eqn:CRMN}
\end{subequations}

Returning to the macroscale equation of motion \eref{TDaveragedisplacement}, we integrate and substitute \eref{cubicquadratic}
\begin{align}
u(x)_{,tt} =& \frac 1 l \int_0^l \frac{(E(y)u(x,y)_{,y})_{,y}}{\rho(y)} {\rm d}y, \notag \\
=& \frac{1}{l(2\alpha \rho_s+\beta \rho_c )} (E(y)u(x,y)_{,y}) \Big\vert^{y = l}_{y=0}, \notag \\
=& \frac{1}{l(2\alpha \rho_s+\beta \rho_c )} E_s(u(x+l,0)_{,y}-u(x,0)_{,y}), \notag \\
=& \frac{1}{l(2\alpha \rho_s+\beta \rho_c )} E_s(a(x+l)-a(x)). 
\label{eqn:43TDaveragedisplacement}
\end{align}
Now substitute \eref{A_sol} and \eref{CRMN} into \eref{43TDaveragedisplacement} and utilize the definition of an inverse Fourier transform to give
\begin{align}
u(x_0)_{,tt} =& \frac{1}{l(2\alpha \rho_s+\beta \rho_c )} E_s(a(x_0+l)-a(x_0)), \notag \\
=& \frac{E_s}{l(2\alpha \rho_s+\beta \rho_c )}\int^{+\infty}_{-\infty}(e^{i2\pi l\xi}-1)A(\xi)e^{i2\pi\xi x_0} {\rm d}\xi \notag \\
=& \frac{E_s}{l(2\alpha \rho_s+\beta \rho_c )}\int^{+\infty}_{-\infty} \frac{12(e^{i2\pi l\xi}-1)^2( 2\alpha \frac{\rho_s}{\rho_c}+\beta) U(\xi)e^{i2\pi\xi x_0} }{l^2( a_0  +a_1e^{i2\pi l\xi} +a_2e^{i4\pi l\xi} )   }{\rm d}\xi \notag \\
= & \frac{E_s}{\rho_c}\int^{+\infty}_{-\infty} \frac{12(e^{i2\pi l\xi}-1)^2  e^{i2\pi\xi x_0}  U(\xi)}{l^2( a_0  +a_1e^{i2\pi l\xi} +a_2e^{i4\pi l\xi} )}{\rm d}\xi .
\label{eqn:Expressionforux0}
\end{align}
The Fourier transform of $u(x)$ is
\begin{align}
    U(\xi) = \int^{+\infty}_{-\infty} u(x)e^{-i2\pi\xi x} {\rm d}x. 
    \label{eqn:FTaveragedisplacement}
\end{align}
Substituting \eref{FTaveragedisplacement} into \eref{Expressionforux0} gives
\begin{align}
    u(x_0)_{,tt} = & \frac{E_s}{\rho_c} \int^{+\infty}_{-\infty}\int^{+\infty}_{-\infty} \frac{12(e^{i2\pi l\xi}-1)^2  u(x) e^{i2\pi\xi (x_0-x)} }{ l^2\left(a_0+a_1e^{i2\pi l\xi} +a_2e^{i4\pi l\xi} \right)  }{\rm d}x {\rm d}\xi,  \notag \\
= & \frac{E_s}{\rho_c} \int^{+\infty}_{-\infty} \left( \int^{+\infty}_{-\infty} \frac{12(e^{i2\pi l\xi}-1)^2  e^{i2\pi\xi (x_0-x)} }{l^2( a_0  +a_1 e^{i2\pi l\xi} +a_2 e^{i4\pi l\xi} )  } {\rm d}\xi  \right)  u(x) {\rm d}x ,   \notag \\
= & \frac{E_s}{\rho_c} \int^{+\infty}_{-\infty} \omega(x_0-x) u(x) {\rm d}x   . 
\label{eqn:nonlocaldynamicequation}
\end{align}
where $\frac{E_s}{\rho_c} \omega(x_0-x)$ will be the \emph{kernel function} of the model and the \emph{influence fucntion} $\omega(x_0-x)$ is defined as
\begin{align}
    \omega(x_0-x) =  \int^{+\infty}_{-\infty}  \frac{12(e^{i2\pi l\xi}-1)^2  e^{i2\pi\xi (x_0-x)} }{l^2( a_0  +a_1 e^{i2\pi l\xi} +a_2 e^{i4\pi l\xi} )  } {\rm d}\xi,
    \label{eqn:nonlocalkernel}
\end{align}
which is the inverse Fourier transform of 
\[
    \Omega(\xi) = \frac{12(e^{i2\pi l\xi}-1)^2 }{l^2( a_0  +a_1 e^{i2\pi l\xi} +a_2 e^{i4\pi l\xi} )  }.
\]
Notice that $\Omega(\xi)$ is a periodic function with period $\nicefrac{1}{l}$, therefore it can be written as Fourier series
\begin{align}
    \Omega(\xi) = \sum^{+\infty}_{n=-\infty} c_n e^{i2n\pi l\xi},
    \label{eqn:FourierseriesofOmega}
\end{align}
where 
\begin{align}
    c_n = l \int^{\frac 1 l}_0 \Omega(\xi) e^{-i2n\pi l\xi} {\rm d}\xi.
    \label{eqn:discretekernel}
\end{align}
Substituting \eref{FourierseriesofOmega} into \eref{nonlocalkernel} gives
\begin{align}
\omega(x_0-x) =& \int^{+\infty}_{-\infty} \Omega (\xi) e^{i2\pi\xi (x_0-x)} {\rm d}\xi, \notag \\
=& \sum^{+\infty}_{n=-\infty} c_n \int^{+\infty}_{-\infty}  e^{i2\pi\xi (x_0+nl-x)} {\rm d} \xi ,\notag \\
=& \sum^{+\infty}_{n=-\infty} c_n \delta(x_0+nl-x).
\label{eqn:discretenonlocalkernel}
\end{align}
where $\delta$ is the Dirac delta function. The kernel $\omega(x_0-x)$ turns out to be a discrete kernel and the discretization length scale is $l$ coinciding with the microscopic length scale. This makes sense because the macroscopic displacement \eref{averagedisplacement} is defined by taking the average displacement of every $l$ interval. We can further simplify \eref{nonlocaldynamicequation} using the notation of Fourier transform ($\mathcal{F}$) and the convolution operator ($*$)
\begin{align}
    u(x_0)_{,tt} =& \frac{E_s}{\rho_c} \int^{+\infty}_{-\infty} \omega(x_0-x)u(x) {\rm d}x ,\notag \\
    =& \frac{E_s}{\rho_c} ((\mathcal{F}^{-1}\Omega)*u)(x_0),
    \label{eqn:FToftt}
\end{align}
where $\mathcal{F}^{-1}$ is the inverse Fourier transform operator. Performing Fourier transform on both side of \eref{FToftt} gives
\begin{align}
    \mathcal{F}(u(x_0)_{,tt})(\xi) =& \frac{E_s}{\rho_c} \mathcal{F}(\mathcal F^{-1}\Omega)(\xi) \cdot \mathcal{F}u(\xi) ,\notag \\
    =& \frac{E_s}{\rho_c} \Omega(\xi) \mathcal{F}u(\xi).
    \label{eqn:FTFToftt}
\end{align}
If we allow the length of unit cell $l \to 0$ then $\Omega(\xi)$ will converge as shown
\begin{align}
    \lim_{l\to 0}\Omega(\xi) =& \lim_{l\to 0}\frac{12(e^{i2\pi l\xi}-1)^2 }{l^2( a_0  +a_1 e^{i2\pi l\xi} +a_2 e^{i4\pi l\xi} )  } ,\notag \\
    =& \frac{12(-4\pi^2\xi^2)}{a_0+a_1+a_2}  .
    \label{eqn:tempft}
\end{align}
Substitute \eref{CRMN} into \eref{tempft} and simplify the equation with $2\alpha + \beta = 1$, we have
\begin{align}
    \lim_{l\to 0}\Omega(\xi)=& \frac{-4\pi^2\xi^2}{(\beta\frac{E_s}{E_c}+2\alpha)(2\alpha\frac{\rho_s}{\rho_c}+\beta)} .
    \label{eqn:limofOmega}
\end{align}
Therefore, with \eref{FToftt} and \eref{limofOmega} we can evaluate the limit of $u(x_0)_{,tt}$ with a Fourier transform
\begin{align}
    \lim_{l\to 0}\mathcal{F}(u(x_0)_{,tt})(\xi)   =& \lim_{l\to 0}\frac{E_s}{\rho_c} \Omega(\xi) \mathcal{F}u(\xi), \notag \\
    =& \frac{E_s}{\rho_c}\frac{-4\pi^2\xi^2}{(\beta\frac{E_s}{E_c}+2\alpha)(2\alpha\frac{\rho_s}{\rho_c}+\beta)} \mathcal{F}u(\xi),  \notag \\
    =& \frac{-4\pi^2\xi^2}{\left(\frac{2\alpha}{E_s}+\frac{\beta}{E_c}\right)\left(2\alpha\rho_s+\beta\rho_c \right)} \mathcal{F}u(\xi), \notag \\
    =& \frac{1}{\left(\frac{2\alpha}{E_s}+\frac{\beta}{E_c}\right)\left(2\alpha\rho_s+\beta\rho_c \right)} u(x_0)_{,xx},
\end{align}
and because the Fourier transform operator is continuous, we can conclude that
\begin{align}
    \lim_{l\to 0}u(x_0)_{,tt} = \frac{E_{ave}}{\rho_{ave}}u(x_0)_{,xx},
\end{align}
where $E_{ave}$ and $\rho_{ave}$ are the homogenized elastic modulus and density
\[ E_{ave} = \frac{1}{\frac{2\alpha}{E_s}+\frac{\beta}{E_c}}, \qquad \rho_{ave} = 2\alpha \rho_s+\beta\rho_c,\]
which demonstrates consistency of our nonlocal model and the classical homogenization theory \citep{murakami1981mixture}.

    Now we will make a few remarks about important properties of the discrete influence function $c_n$.
\begin{remark}
$c_n$ is real and $c_n = c_{-n}$ for all $n\in \mathbb{Z}$. 

Notice that \eref{CRMN} shows that $a_0 = a_2$, so $c_n$ can be written as
\begin{align}
    c_n =&  \int^{\frac 1 l}_0 \Omega(\xi) e^{-i2n\pi l\xi} {\rm d}\xi, \notag \\
    =& \int^{\frac 1 l}_0    \frac{12(e^{i2\pi l\xi}-1)^2 }{l( a_0  +a_1 e^{i2\pi l\xi} +a_2 e^{i4\pi l\xi} )  }  e^{-i2n\pi l\xi} {\rm d}\xi, \notag \\
    =&\frac{12}{l} \int^{\frac 1 l}_0    \frac{e^{i2\pi l\xi}-2+e^{-i2\pi l\xi}  }{ a_0 e^{-i2\pi l\xi}  +a_1 +a_2 e^{i2\pi l\xi}   }  e^{-i2n\pi l\xi} {\rm d}\xi, \notag \\
    =&\frac{12}{l} \int^{\frac 1 l}_0    \frac{2\cos(2\pi l\xi)-2  }{ 2a_0 \cos(2\pi l\xi)  +a_1 }  e^{-i2n\pi l\xi} {\rm d}\xi. 
    \label{eqn:discretekernel}
\end{align}
Since $\frac{2\cos(2\pi l\xi)-2  }{ 2a_0 \cos(2\pi l\xi)  +a_1 }$ is symmetric for $\xi \in [0,\frac 1 l]$, it is straightforward to verify that $c_n \in \mathbb R$ and $c_n = c_{-n}$ for $\forall n \in \mathbb Z$.
\end{remark}

\begin{remark}
$\vert c_n\vert$ is decaying exponentially.

Define complex function $F(z)$ as
\[F(z) = \frac{(24\alpha + 12\beta)(z-1)^2 }{l( a_0  +a_1 z +a_2 z^2 )  }, \qquad \Omega(\xi) = F\left(e^{i2\pi l\xi}\right). \]
If there exist $\epsilon > 0$ and $F(z)$ is analytical on $1-\epsilon \le \vert z\vert \le 1+\epsilon$, then consider the Fourier coefficients of $G(\xi) := F\left((1+\epsilon)e^{i2\pi l\xi}\right)$
\begin{align}
    \int^{\frac 1 l}_0 G(\xi)e^{-i 2n\pi l\xi} {\rm d}\xi &= \int^{\frac 1 l}_0 F((1+\epsilon)e^{i2\pi l\xi})e^{-i 2n\pi l\xi} {\rm d}\xi,  \notag \\
    &= (1+\epsilon)^{n}\oint_{\vert z \vert = 1+\epsilon} F(z) z^{-n} \frac{ {\rm d} z} {i2\pi lz}, \notag \\
    &= (1+\epsilon)^n c_n, \label{eqn:cauchy}
\end{align}
where the Cauchy integral theorem has been used in the last step leading to \eref{cauchy}. The Riemann-Lebesgue lemma requires that Fourier coefficients of vanish at infinity, so we have $(1+\epsilon)^n c_n \to 0$ as $n \to \infty$. Restated, $\vert c_n\vert$ is decaying exponentially, which means we can accurately evaluate the summation \eref{discretenonlocalkernel} by truncation at a finite $n$.
\end{remark}
\begin{remark}
$\sum_{n=-\infty}^{+\infty} c_n=0$.

Let $\xi = 0$ in \eref{nonlocalkernel}, then
\begin{align}
    \sum^{+\infty}_{-\infty} c_n &= \Omega(0) = 0.
\end{align}
Therefore, after we substitute \eref{discretenonlocalkernel} into \eref{nonlocaldynamicequation}, we can express the equation for macroscopic displacement as
\begin{align}
    u(x_0)_{,tt} =& \frac{E_s}{\rho_c} \int^{+\infty}_{-\infty}\sum^{+\infty}_{n=-\infty}c_n \delta(x_0+nl-x)u(x) {\rm d} x, \notag \\
    =& \frac{E_s}{\rho_c} \sum^{+\infty}_{n=-\infty}c_n u(x_0+nl) ,\notag \\
    =& \frac{E_s}{\rho_c} \sum^{+\infty}_{n=-\infty}c_n \left(u(x_0+nl)-u(x_0)\right) ,
\end{align}
which is consistent with the Riemann discretization of the peridynamic intergral with $c_n$ being the discrete peridynamic influence function and $\frac{E_s}{\rho_c}c_n$ being the discrete peridynamic kernel function.
\end{remark}

\section{Higher Order Models}
\label{chap:higher}
The highest order of displacement continuity at interfaces used in the derivation of the last section \eref{extracontinuity} is in the form
\begin{align}
    \frac{E(y^-)}{\rho(y^-)}u(x,y^-)_{,yy} = \frac{E(y^+)}{\rho(y^+)}u(x,y^+)_{,yy}
\end{align}
so we call the nonlocal kernel derived in the previous section the \emph{second order nonlocal kernel}. Any higher order continuity equation at interface can be used to derive the governing equation \eref{governingequations}; therefore, we use higher order polynomials to approximate the displacement inside the unit cell and derive a \emph{fourth-order nonlocal kernel}.

In addition to the continuity equations \eref{continuity1} we add third- and fourth-order continuity equations.  The third-order equations are 
\begin{align*}
    E(y^-) u(x,y^-)_{,y} &= E(y^+) u(x,y^+)_{,y},  \\
    (E(y^-) u(x,y^-)_{,y})_{,tt} &= (E(y^+) u(x,y^+)_{,y})_{,tt},  \\
    E(y^-) u(x,y^-)_{,tty} &= E(y^+) u(x,y^+)_{,tty}, \\
    E(y^-) \left(\frac{E(y^-)}{\rho(y^-)} u(x,y^-)_{yy}\right)_{,y} &= E(y^+)\left(\frac{E(y^+)}{\rho(y^+)} u(x,y+)_{yy}\right)_{,y}, \\
    E(y^-) \left(\frac{E(y^-)}{\rho(y^-)}\right) u(x,y^-)_{,yyy} &= E(y^+)\left(\frac{E(y^+)}{\rho(y^+)}\right) u(x,y^+)_{,yyy},  
\end{align*}
and the fourth-order equations are
\begin{align*}
    u(x,y^-)_{,tttt} &= u(x,y^-)_{,tttt}, \\
    \left(\frac{E(y^-)}{\rho(y^-)}u(x,y^-)_{,yy}\right)_{,tt} &= \left(\frac{E(y^+)}{\rho(y^+)}u(x,y^+)_{,yy}\right)_{,tt}, \\
    \frac{E(y^-)}{\rho(y^-)}u(x,y^-)_{,ttyy}&= \frac{E(y^+)}{\rho(y^+)}u(x,y^+)_{,ttyy}, \\
    \frac{E(y^-)}{\rho(y^-)}\left(\frac{E(y^-)}{\rho(y^-)}u(x,y^-)_{,yy}\right)_{,yy}&= \frac{E(y^+)}{\rho(y^+)}\left(\frac{E(y^+)}{\rho(y^+)}u(x,y^+)_{,yy}\right)_{,yy}, \\
    \left(\frac{E(y^-)}{\rho(y^-)}\right)^{2}u(x,y^-)_{,yyyy}&= \left(\frac{E(y^+)}{\rho(y^+)}\right)^{2}u(x,y^+)_{,yyyy},
\end{align*}
Following arguments leading to \eref{extracontinuity} there will be six more equations inside the unit cell. Therefore, we can use fourth- and fifth-order polynomial anzatzes to approximate the displacement
\begin{align*}
u(x,y) = 
\begin{cases}
    b_4(x)y^4 + b_3(x)y^3 + b_2(x)y^2 \\
    \quad + b_1(x)y +b_0(x) & 0<y<\alpha l, \\
f_5(x)(y-\alpha l )^5 + f_4(x)(y-\alpha l)^4 \\
\quad + f_3(x)(y-\alpha l)^3   + f_2(x)(y - \alpha l)^2 \\
\quad + f_1(x)(y-\alpha l) + f_0(x) & \alpha l < y< (\alpha + \beta)l, \\
d_4(x)y^4 + d_3(x)(y-\alpha l - \beta l)^3 \\
\quad + d_2(x)(y-\alpha l-\beta l)^2 \\
\quad + d_1(x)(y-\alpha l-\beta l) +d_0(x) & (\alpha+\beta)l<y<(2\alpha+\beta)l.
\end{cases}
\end{align*}
We then proceed with the rest of the analysis following exactly as in the previous section using computer symbolic algebraic manipulation.  In the interest of brevity, we will not show the rather lengthy final form of the equations; however, we will demonstrate the accuracy of the theory with numerical experiments in the next section.

%% file: Numerical.tex
\section{Numerical Example}
\label{chap:numerical}
\begin{figure}[h]
\centering
\scalebox{0.6}{\input{discrete_kernel_comparison.pgf}}
\caption{$c_n$ for various polynomial ansatzes}
\label{fig:kernel_func}
\end{figure}
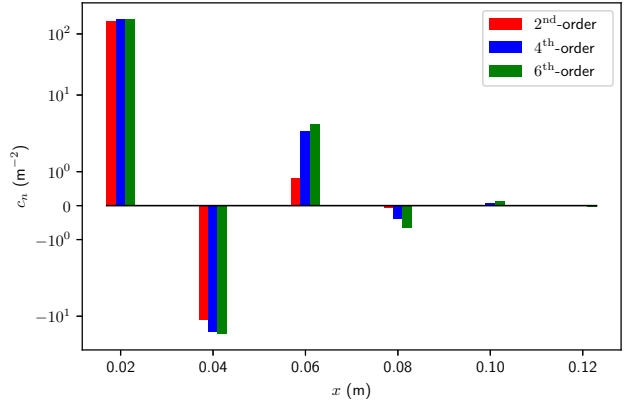

In this section, we will conduct several numerical experiments on the problem described in {\S\ref{chap:problem}} to demonstrate the accuracy of the derived nonlocal kernels in numerical simulation.  The geometric values used in our numerical experiments are as follows: $L = 1 \textrm{m}$, area cross section $A = 10^{-4} \textrm{m}^2$, $l = 0.02 \textrm{m}$. The material properties values are: $E_h = 200 \textrm{GPa}$, $E_s = 5 \textrm{GPa}$; $\rho_h = \rho_s = 8000 \textrm{kg/m}^3$. The bar is subjected to a time-dependent displacement boundary condition $u_{bc}(t) = u_0 a_0 t^6(t-T)^6[1-H(t-T)]$, where $u_0 = -5\times10^{-5} \mathrm{m}$, $a_0$ is a scaling factor, $H$ is the Heaviside function and $T = 0.157 \textrm{ms}$. Substituting these parameters into \eref{discretenonlocalkernel} for the second-order nonlocal influence function and evaluating, results in the values shown in Table~\ref{table:cn1-4}.
\begin{table}[ht]
\caption{$2^{\mathrm{nd}}$ order discrete influence function}
\centering
\begin{tabular}{c c c c c c c}
\hline
$n$ &\vline & 1 & 2 & 3 & 4 & $\ge$ 5 \\
\hline
$c_n$ &\vline & 161.3418 & -11.4752 & 0.8162 & -0.058 & $\vert c_n\vert <$ 0.01 \\
\hline
\end{tabular}
\label{table:cn1-4}
\end{table}
As indicated, the discrete influence function decreases rapidly with increasing $n$. Therefore, we truncate the summation in \eref{discretenonlocalkernel} and only use the terms of $\vert n \vert\le 6$, which is the equivalent to setting the horizon to be $\epsilon = 6l$. In this regard, the horizon size of the nonlocal kernel as well as the discrete node-spacing is completely determined by material's microstructure and the multiscale model. We computed fourth-order and sixth-order kernels as well and they are shown graphically in Figure~\ref{fig:kernel_func}.

For comparison we also use a standard peridynamic kernel to solve the displacement of the bar using the local homogenized material properties
\begin{align}
    E_{ave} = \frac{1}{\frac{2\alpha}{E_h} + \frac{\beta}{E_s}}, \qquad    \rho_{ave} = 2\alpha \rho_h+\beta\rho_s.
\end{align}
To be specific, we use the peridynamic equation \citep{silling2005meshfree}
\begin{align}
    \ddot{u}(x) =& \int^{+\epsilon}_{-\epsilon} \omega_p(\vert\xi\vert)\frac{E_{ave}}{\rho_{ave}}\left( u(x + \xi) - u(x) \right) \rm{d} \xi, \notag \\
    \omega_p(\vert \xi&\vert) = \frac{2}{\epsilon^2\vert\xi\vert},
\end{align}
where $\epsilon$ is the peridynamic horizon and $\omega_p(\vert\xi\vert)\frac{E_{ave}}{\rho_{ave}}$ is the corresponding kernel function.

\begin{figure}[h]
\centering
\scalebox{0.6}{\input{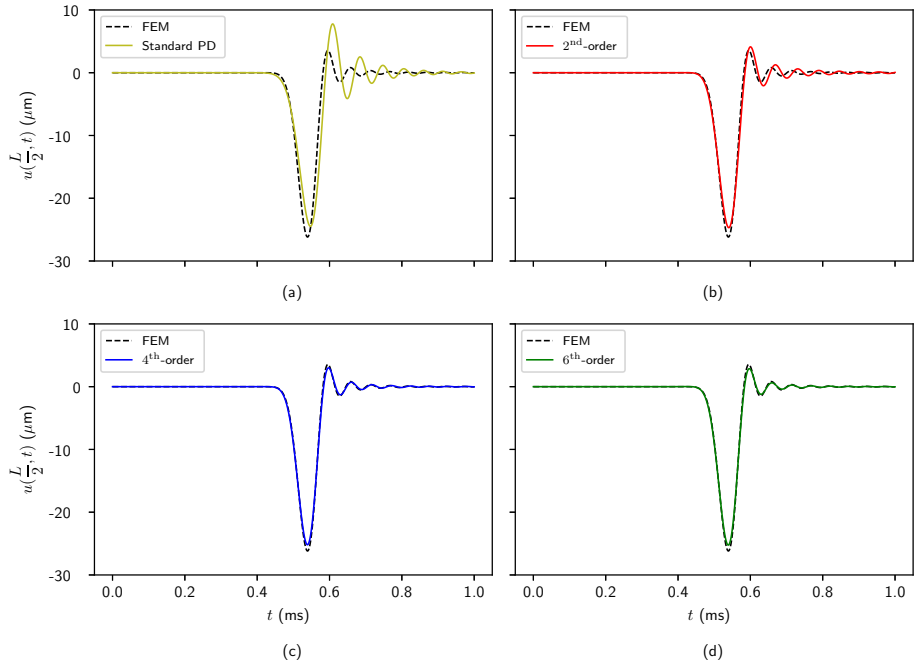}}
\caption{Midpoint displacement comparison of the derived nonlocal kernels and a standard peridynamics kernel}
\label{fig:NonlocalPDFEM}
\end{figure}

With no guidance as to how to choose $\epsilon$ and the discretization node spacing in the standard model, we resort to trial-and-error to achieve the best results when comparing with a reference solution of a highly-resolved microstructural finite element model (FEM) that accurately captures the wave dynamics of the bar.  After many attempts, the peridynamic node spacing is set to be $l_p=0.005$m and horizon size is $\epsilon = 4l_p$.  Perhaps we could achieve better results by using an optimization framework to select the parameters, but no choice in our trials gave near-accurate results. Of course, there are infinite choices of $\omega_p$ we could have investigated as well; our choice here reflects the one of the most common choices found in the literature.  We present midpoint displacements for the standard peridynamics kernel and our microstructural derived kernels when compared with the FEM  reference solution in Figure~\ref{fig:NonlocalPDFEM}.
It is shown that the microstructural derived kernels are more accurate that the standard peridynamics kernel.  Additionally, the accuracy improves with increasing order of the polynomial anatz; however, the difference between the forth-order and the sixth-order derived kernels appears to be negligible.

Regarding the computation complexity, the nonlocal kernel methods only assign one peridynamic node to each unit cell, while the classical FEM needs several elements per each constituent inside unit cell in order to capture the wave dynamics accurately.  Normally, a nonlocal method is computationally slower that a local method with an equivalent discretization length scale due to the need to integrate the nonlocal interactions; however, in this case, the dispersive nature of the nonlocal model along with the derived kernel gives good results at a lower computational cost than the fully resolved local model.

%% file: discrete_kernel_comparison.pgf
\begingroup%
\makeatletter%
\begin{pgfpicture}%
\pgfpathrectangle{\pgfpointorigin}{\pgfqpoint{6.000000in}{4.000000in}}%
\pgfusepath{use as bounding box, clip}%
\begin{pgfscope}%
\pgfsetbuttcap%
\pgfsetmiterjoin%
\definecolor{currentfill}{rgb}{1.000000,1.000000,1.000000}%
\pgfsetfillcolor{currentfill}%
\pgfsetlinewidth{0.000000pt}%
\definecolor{currentstroke}{rgb}{1.000000,1.000000,1.000000}%
\pgfsetstrokecolor{currentstroke}%
\pgfsetdash{}{0pt}%
\pgfpathmoveto{\pgfqpoint{0.000000in}{0.000000in}}%
\pgfpathlineto{\pgfqpoint{6.000000in}{0.000000in}}%
\pgfpathlineto{\pgfqpoint{6.000000in}{4.000000in}}%
\pgfpathlineto{\pgfqpoint{0.000000in}{4.000000in}}%
\pgfpathclose%
\pgfusepath{fill}%
\end{pgfscope}%
\begin{pgfscope}%
\pgfsetbuttcap%
\pgfsetmiterjoin%
\definecolor{currentfill}{rgb}{1.000000,1.000000,1.000000}%
\pgfsetfillcolor{currentfill}%
\pgfsetlinewidth{0.000000pt}%
\definecolor{currentstroke}{rgb}{0.000000,0.000000,0.000000}%
\pgfsetstrokecolor{currentstroke}%
\pgfsetstrokeopacity{0.000000}%
\pgfsetdash{}{0pt}%
\pgfpathmoveto{\pgfqpoint{0.750000in}{0.500000in}}%
\pgfpathlineto{\pgfqpoint{5.400000in}{0.500000in}}%
\pgfpathlineto{\pgfqpoint{5.400000in}{3.520000in}}%
\pgfpathlineto{\pgfqpoint{0.750000in}{3.520000in}}%
\pgfpathclose%
\pgfusepath{fill}%
\end{pgfscope}%
\begin{pgfscope}%
\pgfpathrectangle{\pgfqpoint{0.750000in}{0.500000in}}{\pgfqpoint{4.650000in}{3.020000in}}%
\pgfusepath{clip}%
\pgfsetbuttcap%
\pgfsetmiterjoin%
\definecolor{currentfill}{rgb}{1.000000,0.000000,0.000000}%
\pgfsetfillcolor{currentfill}%
\pgfsetlinewidth{0.000000pt}%
\definecolor{currentstroke}{rgb}{0.000000,0.000000,0.000000}%
\pgfsetstrokecolor{currentstroke}%
\pgfsetstrokeopacity{0.000000}%
\pgfsetdash{}{0pt}%
\pgfpathmoveto{\pgfqpoint{0.961364in}{1.755635in}}%
\pgfpathlineto{\pgfqpoint{1.041123in}{1.755635in}}%
\pgfpathlineto{\pgfqpoint{1.041123in}{3.360035in}}%
\pgfpathlineto{\pgfqpoint{0.961364in}{3.360035in}}%
\pgfpathclose%
\pgfusepath{fill}%
\end{pgfscope}%
\begin{pgfscope}%
\pgfpathrectangle{\pgfqpoint{0.750000in}{0.500000in}}{\pgfqpoint{4.650000in}{3.020000in}}%
\pgfusepath{clip}%
\pgfsetbuttcap%
\pgfsetmiterjoin%
\definecolor{currentfill}{rgb}{1.000000,0.000000,0.000000}%
\pgfsetfillcolor{currentfill}%
\pgfsetlinewidth{0.000000pt}%
\definecolor{currentstroke}{rgb}{0.000000,0.000000,0.000000}%
\pgfsetstrokecolor{currentstroke}%
\pgfsetstrokeopacity{0.000000}%
\pgfsetdash{}{0pt}%
\pgfpathmoveto{\pgfqpoint{1.758962in}{1.755635in}}%
\pgfpathlineto{\pgfqpoint{1.838722in}{1.755635in}}%
\pgfpathlineto{\pgfqpoint{1.838722in}{0.761552in}}%
\pgfpathlineto{\pgfqpoint{1.758962in}{0.761552in}}%
\pgfpathclose%
\pgfusepath{fill}%
\end{pgfscope}%
\begin{pgfscope}%
\pgfpathrectangle{\pgfqpoint{0.750000in}{0.500000in}}{\pgfqpoint{4.650000in}{3.020000in}}%
\pgfusepath{clip}%
\pgfsetbuttcap%
\pgfsetmiterjoin%
\definecolor{currentfill}{rgb}{1.000000,0.000000,0.000000}%
\pgfsetfillcolor{currentfill}%
\pgfsetlinewidth{0.000000pt}%
\definecolor{currentstroke}{rgb}{0.000000,0.000000,0.000000}%
\pgfsetstrokecolor{currentstroke}%
\pgfsetstrokeopacity{0.000000}%
\pgfsetdash{}{0pt}%
\pgfpathmoveto{\pgfqpoint{2.556561in}{1.755635in}}%
\pgfpathlineto{\pgfqpoint{2.636321in}{1.755635in}}%
\pgfpathlineto{\pgfqpoint{2.636321in}{1.996693in}}%
\pgfpathlineto{\pgfqpoint{2.556561in}{1.996693in}}%
\pgfpathclose%
\pgfusepath{fill}%
\end{pgfscope}%
\begin{pgfscope}%
\pgfpathrectangle{\pgfqpoint{0.750000in}{0.500000in}}{\pgfqpoint{4.650000in}{3.020000in}}%
\pgfusepath{clip}%
\pgfsetbuttcap%
\pgfsetmiterjoin%
\definecolor{currentfill}{rgb}{1.000000,0.000000,0.000000}%
\pgfsetfillcolor{currentfill}%
\pgfsetlinewidth{0.000000pt}%
\definecolor{currentstroke}{rgb}{0.000000,0.000000,0.000000}%
\pgfsetstrokecolor{currentstroke}%
\pgfsetstrokeopacity{0.000000}%
\pgfsetdash{}{0pt}%
\pgfpathmoveto{\pgfqpoint{3.354160in}{1.755635in}}%
\pgfpathlineto{\pgfqpoint{3.433919in}{1.755635in}}%
\pgfpathlineto{\pgfqpoint{3.433919in}{1.738491in}}%
\pgfpathlineto{\pgfqpoint{3.354160in}{1.738491in}}%
\pgfpathclose%
\pgfusepath{fill}%
\end{pgfscope}%
\begin{pgfscope}%
\pgfpathrectangle{\pgfqpoint{0.750000in}{0.500000in}}{\pgfqpoint{4.650000in}{3.020000in}}%
\pgfusepath{clip}%
\pgfsetbuttcap%
\pgfsetmiterjoin%
\definecolor{currentfill}{rgb}{1.000000,0.000000,0.000000}%
\pgfsetfillcolor{currentfill}%
\pgfsetlinewidth{0.000000pt}%
\definecolor{currentstroke}{rgb}{0.000000,0.000000,0.000000}%
\pgfsetstrokecolor{currentstroke}%
\pgfsetstrokeopacity{0.000000}%
\pgfsetdash{}{0pt}%
\pgfpathmoveto{\pgfqpoint{4.151758in}{1.755635in}}%
\pgfpathlineto{\pgfqpoint{4.231518in}{1.755635in}}%
\pgfpathlineto{\pgfqpoint{4.231518in}{1.756855in}}%
\pgfpathlineto{\pgfqpoint{4.151758in}{1.756855in}}%
\pgfpathclose%
\pgfusepath{fill}%
\end{pgfscope}%
\begin{pgfscope}%
\pgfpathrectangle{\pgfqpoint{0.750000in}{0.500000in}}{\pgfqpoint{4.650000in}{3.020000in}}%
\pgfusepath{clip}%
\pgfsetbuttcap%
\pgfsetmiterjoin%
\definecolor{currentfill}{rgb}{1.000000,0.000000,0.000000}%
\pgfsetfillcolor{currentfill}%
\pgfsetlinewidth{0.000000pt}%
\definecolor{currentstroke}{rgb}{0.000000,0.000000,0.000000}%
\pgfsetstrokecolor{currentstroke}%
\pgfsetstrokeopacity{0.000000}%
\pgfsetdash{}{0pt}%
\pgfpathmoveto{\pgfqpoint{4.949357in}{1.755635in}}%
\pgfpathlineto{\pgfqpoint{5.029117in}{1.755635in}}%
\pgfpathlineto{\pgfqpoint{5.029117in}{1.755549in}}%
\pgfpathlineto{\pgfqpoint{4.949357in}{1.755549in}}%
\pgfpathclose%
\pgfusepath{fill}%
\end{pgfscope}%
\begin{pgfscope}%
\pgfpathrectangle{\pgfqpoint{0.750000in}{0.500000in}}{\pgfqpoint{4.650000in}{3.020000in}}%
\pgfusepath{clip}%
\pgfsetbuttcap%
\pgfsetmiterjoin%
\definecolor{currentfill}{rgb}{0.000000,0.000000,1.000000}%
\pgfsetfillcolor{currentfill}%
\pgfsetlinewidth{0.000000pt}%
\definecolor{currentstroke}{rgb}{0.000000,0.000000,0.000000}%
\pgfsetstrokecolor{currentstroke}%
\pgfsetstrokeopacity{0.000000}%
\pgfsetdash{}{0pt}%
\pgfpathmoveto{\pgfqpoint{1.041123in}{1.755635in}}%
\pgfpathlineto{\pgfqpoint{1.120883in}{1.755635in}}%
\pgfpathlineto{\pgfqpoint{1.120883in}{3.378849in}}%
\pgfpathlineto{\pgfqpoint{1.041123in}{3.378849in}}%
\pgfpathclose%
\pgfusepath{fill}%
\end{pgfscope}%
\begin{pgfscope}%
\pgfpathrectangle{\pgfqpoint{0.750000in}{0.500000in}}{\pgfqpoint{4.650000in}{3.020000in}}%
\pgfusepath{clip}%
\pgfsetbuttcap%
\pgfsetmiterjoin%
\definecolor{currentfill}{rgb}{0.000000,0.000000,1.000000}%
\pgfsetfillcolor{currentfill}%
\pgfsetlinewidth{0.000000pt}%
\definecolor{currentstroke}{rgb}{0.000000,0.000000,0.000000}%
\pgfsetstrokecolor{currentstroke}%
\pgfsetstrokeopacity{0.000000}%
\pgfsetdash{}{0pt}%
\pgfpathmoveto{\pgfqpoint{1.838722in}{1.755635in}}%
\pgfpathlineto{\pgfqpoint{1.918482in}{1.755635in}}%
\pgfpathlineto{\pgfqpoint{1.918482in}{0.659356in}}%
\pgfpathlineto{\pgfqpoint{1.838722in}{0.659356in}}%
\pgfpathclose%
\pgfusepath{fill}%
\end{pgfscope}%
\begin{pgfscope}%
\pgfpathrectangle{\pgfqpoint{0.750000in}{0.500000in}}{\pgfqpoint{4.650000in}{3.020000in}}%
\pgfusepath{clip}%
\pgfsetbuttcap%
\pgfsetmiterjoin%
\definecolor{currentfill}{rgb}{0.000000,0.000000,1.000000}%
\pgfsetfillcolor{currentfill}%
\pgfsetlinewidth{0.000000pt}%
\definecolor{currentstroke}{rgb}{0.000000,0.000000,0.000000}%
\pgfsetstrokecolor{currentstroke}%
\pgfsetstrokeopacity{0.000000}%
\pgfsetdash{}{0pt}%
\pgfpathmoveto{\pgfqpoint{2.636321in}{1.755635in}}%
\pgfpathlineto{\pgfqpoint{2.716081in}{1.755635in}}%
\pgfpathlineto{\pgfqpoint{2.716081in}{2.407469in}}%
\pgfpathlineto{\pgfqpoint{2.636321in}{2.407469in}}%
\pgfpathclose%
\pgfusepath{fill}%
\end{pgfscope}%
\begin{pgfscope}%
\pgfpathrectangle{\pgfqpoint{0.750000in}{0.500000in}}{\pgfqpoint{4.650000in}{3.020000in}}%
\pgfusepath{clip}%
\pgfsetbuttcap%
\pgfsetmiterjoin%
\definecolor{currentfill}{rgb}{0.000000,0.000000,1.000000}%
\pgfsetfillcolor{currentfill}%
\pgfsetlinewidth{0.000000pt}%
\definecolor{currentstroke}{rgb}{0.000000,0.000000,0.000000}%
\pgfsetstrokecolor{currentstroke}%
\pgfsetstrokeopacity{0.000000}%
\pgfsetdash{}{0pt}%
\pgfpathmoveto{\pgfqpoint{3.433919in}{1.755635in}}%
\pgfpathlineto{\pgfqpoint{3.513679in}{1.755635in}}%
\pgfpathlineto{\pgfqpoint{3.513679in}{1.639109in}}%
\pgfpathlineto{\pgfqpoint{3.433919in}{1.639109in}}%
\pgfpathclose%
\pgfusepath{fill}%
\end{pgfscope}%
\begin{pgfscope}%
\pgfpathrectangle{\pgfqpoint{0.750000in}{0.500000in}}{\pgfqpoint{4.650000in}{3.020000in}}%
\pgfusepath{clip}%
\pgfsetbuttcap%
\pgfsetmiterjoin%
\definecolor{currentfill}{rgb}{0.000000,0.000000,1.000000}%
\pgfsetfillcolor{currentfill}%
\pgfsetlinewidth{0.000000pt}%
\definecolor{currentstroke}{rgb}{0.000000,0.000000,0.000000}%
\pgfsetstrokecolor{currentstroke}%
\pgfsetstrokeopacity{0.000000}%
\pgfsetdash{}{0pt}%
\pgfpathmoveto{\pgfqpoint{4.231518in}{1.755635in}}%
\pgfpathlineto{\pgfqpoint{4.311278in}{1.755635in}}%
\pgfpathlineto{\pgfqpoint{4.311278in}{1.773330in}}%
\pgfpathlineto{\pgfqpoint{4.231518in}{1.773330in}}%
\pgfpathclose%
\pgfusepath{fill}%
\end{pgfscope}%
\begin{pgfscope}%
\pgfpathrectangle{\pgfqpoint{0.750000in}{0.500000in}}{\pgfqpoint{4.650000in}{3.020000in}}%
\pgfusepath{clip}%
\pgfsetbuttcap%
\pgfsetmiterjoin%
\definecolor{currentfill}{rgb}{0.000000,0.000000,1.000000}%
\pgfsetfillcolor{currentfill}%
\pgfsetlinewidth{0.000000pt}%
\definecolor{currentstroke}{rgb}{0.000000,0.000000,0.000000}%
\pgfsetstrokecolor{currentstroke}%
\pgfsetstrokeopacity{0.000000}%
\pgfsetdash{}{0pt}%
\pgfpathmoveto{\pgfqpoint{5.029117in}{1.755635in}}%
\pgfpathlineto{\pgfqpoint{5.108877in}{1.755635in}}%
\pgfpathlineto{\pgfqpoint{5.108877in}{1.752948in}}%
\pgfpathlineto{\pgfqpoint{5.029117in}{1.752948in}}%
\pgfpathclose%
\pgfusepath{fill}%
\end{pgfscope}%
\begin{pgfscope}%
\pgfpathrectangle{\pgfqpoint{0.750000in}{0.500000in}}{\pgfqpoint{4.650000in}{3.020000in}}%
\pgfusepath{clip}%
\pgfsetbuttcap%
\pgfsetmiterjoin%
\definecolor{currentfill}{rgb}{0.000000,0.500000,0.000000}%
\pgfsetfillcolor{currentfill}%
\pgfsetlinewidth{0.000000pt}%
\definecolor{currentstroke}{rgb}{0.000000,0.000000,0.000000}%
\pgfsetstrokecolor{currentstroke}%
\pgfsetstrokeopacity{0.000000}%
\pgfsetdash{}{0pt}%
\pgfpathmoveto{\pgfqpoint{1.120883in}{1.755635in}}%
\pgfpathlineto{\pgfqpoint{1.200643in}{1.755635in}}%
\pgfpathlineto{\pgfqpoint{1.200643in}{3.382727in}}%
\pgfpathlineto{\pgfqpoint{1.120883in}{3.382727in}}%
\pgfpathclose%
\pgfusepath{fill}%
\end{pgfscope}%
\begin{pgfscope}%
\pgfpathrectangle{\pgfqpoint{0.750000in}{0.500000in}}{\pgfqpoint{4.650000in}{3.020000in}}%
\pgfusepath{clip}%
\pgfsetbuttcap%
\pgfsetmiterjoin%
\definecolor{currentfill}{rgb}{0.000000,0.500000,0.000000}%
\pgfsetfillcolor{currentfill}%
\pgfsetlinewidth{0.000000pt}%
\definecolor{currentstroke}{rgb}{0.000000,0.000000,0.000000}%
\pgfsetstrokecolor{currentstroke}%
\pgfsetstrokeopacity{0.000000}%
\pgfsetdash{}{0pt}%
\pgfpathmoveto{\pgfqpoint{1.918482in}{1.755635in}}%
\pgfpathlineto{\pgfqpoint{1.998242in}{1.755635in}}%
\pgfpathlineto{\pgfqpoint{1.998242in}{0.637273in}}%
\pgfpathlineto{\pgfqpoint{1.918482in}{0.637273in}}%
\pgfpathclose%
\pgfusepath{fill}%
\end{pgfscope}%
\begin{pgfscope}%
\pgfpathrectangle{\pgfqpoint{0.750000in}{0.500000in}}{\pgfqpoint{4.650000in}{3.020000in}}%
\pgfusepath{clip}%
\pgfsetbuttcap%
\pgfsetmiterjoin%
\definecolor{currentfill}{rgb}{0.000000,0.500000,0.000000}%
\pgfsetfillcolor{currentfill}%
\pgfsetlinewidth{0.000000pt}%
\definecolor{currentstroke}{rgb}{0.000000,0.000000,0.000000}%
\pgfsetstrokecolor{currentstroke}%
\pgfsetstrokeopacity{0.000000}%
\pgfsetdash{}{0pt}%
\pgfpathmoveto{\pgfqpoint{2.716081in}{1.755635in}}%
\pgfpathlineto{\pgfqpoint{2.795840in}{1.755635in}}%
\pgfpathlineto{\pgfqpoint{2.795840in}{2.467949in}}%
\pgfpathlineto{\pgfqpoint{2.716081in}{2.467949in}}%
\pgfpathclose%
\pgfusepath{fill}%
\end{pgfscope}%
\begin{pgfscope}%
\pgfpathrectangle{\pgfqpoint{0.750000in}{0.500000in}}{\pgfqpoint{4.650000in}{3.020000in}}%
\pgfusepath{clip}%
\pgfsetbuttcap%
\pgfsetmiterjoin%
\definecolor{currentfill}{rgb}{0.000000,0.500000,0.000000}%
\pgfsetfillcolor{currentfill}%
\pgfsetlinewidth{0.000000pt}%
\definecolor{currentstroke}{rgb}{0.000000,0.000000,0.000000}%
\pgfsetstrokecolor{currentstroke}%
\pgfsetstrokeopacity{0.000000}%
\pgfsetdash{}{0pt}%
\pgfpathmoveto{\pgfqpoint{3.513679in}{1.755635in}}%
\pgfpathlineto{\pgfqpoint{3.593439in}{1.755635in}}%
\pgfpathlineto{\pgfqpoint{3.593439in}{1.564592in}}%
\pgfpathlineto{\pgfqpoint{3.513679in}{1.564592in}}%
\pgfpathclose%
\pgfusepath{fill}%
\end{pgfscope}%
\begin{pgfscope}%
\pgfpathrectangle{\pgfqpoint{0.750000in}{0.500000in}}{\pgfqpoint{4.650000in}{3.020000in}}%
\pgfusepath{clip}%
\pgfsetbuttcap%
\pgfsetmiterjoin%
\definecolor{currentfill}{rgb}{0.000000,0.500000,0.000000}%
\pgfsetfillcolor{currentfill}%
\pgfsetlinewidth{0.000000pt}%
\definecolor{currentstroke}{rgb}{0.000000,0.000000,0.000000}%
\pgfsetstrokecolor{currentstroke}%
\pgfsetstrokeopacity{0.000000}%
\pgfsetdash{}{0pt}%
\pgfpathmoveto{\pgfqpoint{4.311278in}{1.755635in}}%
\pgfpathlineto{\pgfqpoint{4.391038in}{1.755635in}}%
\pgfpathlineto{\pgfqpoint{4.391038in}{1.792750in}}%
\pgfpathlineto{\pgfqpoint{4.311278in}{1.792750in}}%
\pgfpathclose%
\pgfusepath{fill}%
\end{pgfscope}%
\begin{pgfscope}%
\pgfpathrectangle{\pgfqpoint{0.750000in}{0.500000in}}{\pgfqpoint{4.650000in}{3.020000in}}%
\pgfusepath{clip}%
\pgfsetbuttcap%
\pgfsetmiterjoin%
\definecolor{currentfill}{rgb}{0.000000,0.500000,0.000000}%
\pgfsetfillcolor{currentfill}%
\pgfsetlinewidth{0.000000pt}%
\definecolor{currentstroke}{rgb}{0.000000,0.000000,0.000000}%
\pgfsetstrokecolor{currentstroke}%
\pgfsetstrokeopacity{0.000000}%
\pgfsetdash{}{0pt}%
\pgfpathmoveto{\pgfqpoint{5.108877in}{1.755635in}}%
\pgfpathlineto{\pgfqpoint{5.188636in}{1.755635in}}%
\pgfpathlineto{\pgfqpoint{5.188636in}{1.748406in}}%
\pgfpathlineto{\pgfqpoint{5.108877in}{1.748406in}}%
\pgfpathclose%
\pgfusepath{fill}%
\end{pgfscope}%
\begin{pgfscope}%
\pgfsetbuttcap%
\pgfsetroundjoin%
\definecolor{currentfill}{rgb}{0.000000,0.000000,0.000000}%
\pgfsetfillcolor{currentfill}%
\pgfsetlinewidth{0.803000pt}%
\definecolor{currentstroke}{rgb}{0.000000,0.000000,0.000000}%
\pgfsetstrokecolor{currentstroke}%
\pgfsetdash{}{0pt}%
\pgfsys@defobject{currentmarker}{\pgfqpoint{0.000000in}{-0.048611in}}{\pgfqpoint{0.000000in}{0.000000in}}{%
\pgfpathmoveto{\pgfqpoint{0.000000in}{0.000000in}}%
\pgfpathlineto{\pgfqpoint{0.000000in}{-0.048611in}}%
\pgfusepath{stroke,fill}%
}%
\begin{pgfscope}%
\pgfsys@transformshift{1.081003in}{0.500000in}%
\pgfsys@useobject{currentmarker}{}%
\end{pgfscope}%
\end{pgfscope}%
\begin{pgfscope}%
\definecolor{textcolor}{rgb}{0.000000,0.000000,0.000000}%
\pgfsetstrokecolor{textcolor}%
\pgfsetfillcolor{textcolor}%
\pgftext[x=1.081003in,y=0.402778in,,top]{\color{textcolor}\sffamily\fontsize{10.000000}{12.000000}\selectfont 0.02}%
\end{pgfscope}%
\begin{pgfscope}%
\pgfsetbuttcap%
\pgfsetroundjoin%
\definecolor{currentfill}{rgb}{0.000000,0.000000,0.000000}%
\pgfsetfillcolor{currentfill}%
\pgfsetlinewidth{0.803000pt}%
\definecolor{currentstroke}{rgb}{0.000000,0.000000,0.000000}%
\pgfsetstrokecolor{currentstroke}%
\pgfsetdash{}{0pt}%
\pgfsys@defobject{currentmarker}{\pgfqpoint{0.000000in}{-0.048611in}}{\pgfqpoint{0.000000in}{0.000000in}}{%
\pgfpathmoveto{\pgfqpoint{0.000000in}{0.000000in}}%
\pgfpathlineto{\pgfqpoint{0.000000in}{-0.048611in}}%
\pgfusepath{stroke,fill}%
}%
\begin{pgfscope}%
\pgfsys@transformshift{1.878602in}{0.500000in}%
\pgfsys@useobject{currentmarker}{}%
\end{pgfscope}%
\end{pgfscope}%
\begin{pgfscope}%
\definecolor{textcolor}{rgb}{0.000000,0.000000,0.000000}%
\pgfsetstrokecolor{textcolor}%
\pgfsetfillcolor{textcolor}%
\pgftext[x=1.878602in,y=0.402778in,,top]{\color{textcolor}\sffamily\fontsize{10.000000}{12.000000}\selectfont 0.04}%
\end{pgfscope}%
\begin{pgfscope}%
\pgfsetbuttcap%
\pgfsetroundjoin%
\definecolor{currentfill}{rgb}{0.000000,0.000000,0.000000}%
\pgfsetfillcolor{currentfill}%
\pgfsetlinewidth{0.803000pt}%
\definecolor{currentstroke}{rgb}{0.000000,0.000000,0.000000}%
\pgfsetstrokecolor{currentstroke}%
\pgfsetdash{}{0pt}%
\pgfsys@defobject{currentmarker}{\pgfqpoint{0.000000in}{-0.048611in}}{\pgfqpoint{0.000000in}{0.000000in}}{%
\pgfpathmoveto{\pgfqpoint{0.000000in}{0.000000in}}%
\pgfpathlineto{\pgfqpoint{0.000000in}{-0.048611in}}%
\pgfusepath{stroke,fill}%
}%
\begin{pgfscope}%
\pgfsys@transformshift{2.676201in}{0.500000in}%
\pgfsys@useobject{currentmarker}{}%
\end{pgfscope}%
\end{pgfscope}%
\begin{pgfscope}%
\definecolor{textcolor}{rgb}{0.000000,0.000000,0.000000}%
\pgfsetstrokecolor{textcolor}%
\pgfsetfillcolor{textcolor}%
\pgftext[x=2.676201in,y=0.402778in,,top]{\color{textcolor}\sffamily\fontsize{10.000000}{12.000000}\selectfont 0.06}%
\end{pgfscope}%
\begin{pgfscope}%
\pgfsetbuttcap%
\pgfsetroundjoin%
\definecolor{currentfill}{rgb}{0.000000,0.000000,0.000000}%
\pgfsetfillcolor{currentfill}%
\pgfsetlinewidth{0.803000pt}%
\definecolor{currentstroke}{rgb}{0.000000,0.000000,0.000000}%
\pgfsetstrokecolor{currentstroke}%
\pgfsetdash{}{0pt}%
\pgfsys@defobject{currentmarker}{\pgfqpoint{0.000000in}{-0.048611in}}{\pgfqpoint{0.000000in}{0.000000in}}{%
\pgfpathmoveto{\pgfqpoint{0.000000in}{0.000000in}}%
\pgfpathlineto{\pgfqpoint{0.000000in}{-0.048611in}}%
\pgfusepath{stroke,fill}%
}%
\begin{pgfscope}%
\pgfsys@transformshift{3.473799in}{0.500000in}%
\pgfsys@useobject{currentmarker}{}%
\end{pgfscope}%
\end{pgfscope}%
\begin{pgfscope}%
\definecolor{textcolor}{rgb}{0.000000,0.000000,0.000000}%
\pgfsetstrokecolor{textcolor}%
\pgfsetfillcolor{textcolor}%
\pgftext[x=3.473799in,y=0.402778in,,top]{\color{textcolor}\sffamily\fontsize{10.000000}{12.000000}\selectfont 0.08}%
\end{pgfscope}%
\begin{pgfscope}%
\pgfsetbuttcap%
\pgfsetroundjoin%
\definecolor{currentfill}{rgb}{0.000000,0.000000,0.000000}%
\pgfsetfillcolor{currentfill}%
\pgfsetlinewidth{0.803000pt}%
\definecolor{currentstroke}{rgb}{0.000000,0.000000,0.000000}%
\pgfsetstrokecolor{currentstroke}%
\pgfsetdash{}{0pt}%
\pgfsys@defobject{currentmarker}{\pgfqpoint{0.000000in}{-0.048611in}}{\pgfqpoint{0.000000in}{0.000000in}}{%
\pgfpathmoveto{\pgfqpoint{0.000000in}{0.000000in}}%
\pgfpathlineto{\pgfqpoint{0.000000in}{-0.048611in}}%
\pgfusepath{stroke,fill}%
}%
\begin{pgfscope}%
\pgfsys@transformshift{4.271398in}{0.500000in}%
\pgfsys@useobject{currentmarker}{}%
\end{pgfscope}%
\end{pgfscope}%
\begin{pgfscope}%
\definecolor{textcolor}{rgb}{0.000000,0.000000,0.000000}%
\pgfsetstrokecolor{textcolor}%
\pgfsetfillcolor{textcolor}%
\pgftext[x=4.271398in,y=0.402778in,,top]{\color{textcolor}\sffamily\fontsize{10.000000}{12.000000}\selectfont 0.10}%
\end{pgfscope}%
\begin{pgfscope}%
\pgfsetbuttcap%
\pgfsetroundjoin%
\definecolor{currentfill}{rgb}{0.000000,0.000000,0.000000}%
\pgfsetfillcolor{currentfill}%
\pgfsetlinewidth{0.803000pt}%
\definecolor{currentstroke}{rgb}{0.000000,0.000000,0.000000}%
\pgfsetstrokecolor{currentstroke}%
\pgfsetdash{}{0pt}%
\pgfsys@defobject{currentmarker}{\pgfqpoint{0.000000in}{-0.048611in}}{\pgfqpoint{0.000000in}{0.000000in}}{%
\pgfpathmoveto{\pgfqpoint{0.000000in}{0.000000in}}%
\pgfpathlineto{\pgfqpoint{0.000000in}{-0.048611in}}%
\pgfusepath{stroke,fill}%
}%
\begin{pgfscope}%
\pgfsys@transformshift{5.068997in}{0.500000in}%
\pgfsys@useobject{currentmarker}{}%
\end{pgfscope}%
\end{pgfscope}%
\begin{pgfscope}%
\definecolor{textcolor}{rgb}{0.000000,0.000000,0.000000}%
\pgfsetstrokecolor{textcolor}%
\pgfsetfillcolor{textcolor}%
\pgftext[x=5.068997in,y=0.402778in,,top]{\color{textcolor}\sffamily\fontsize{10.000000}{12.000000}\selectfont 0.12}%
\end{pgfscope}%
\begin{pgfscope}%
\definecolor{textcolor}{rgb}{0.000000,0.000000,0.000000}%
\pgfsetstrokecolor{textcolor}%
\pgfsetfillcolor{textcolor}%
\pgftext[x=3.075000in,y=0.212809in,,top]{\color{textcolor}\sffamily\fontsize{10.000000}{12.000000}\selectfont \(\displaystyle x\) (m)}%
\end{pgfscope}%
\begin{pgfscope}%
\pgfsetbuttcap%
\pgfsetroundjoin%
\definecolor{currentfill}{rgb}{0.000000,0.000000,0.000000}%
\pgfsetfillcolor{currentfill}%
\pgfsetlinewidth{0.803000pt}%
\definecolor{currentstroke}{rgb}{0.000000,0.000000,0.000000}%
\pgfsetstrokecolor{currentstroke}%
\pgfsetdash{}{0pt}%
\pgfsys@defobject{currentmarker}{\pgfqpoint{-0.048611in}{0.000000in}}{\pgfqpoint{0.000000in}{0.000000in}}{%
\pgfpathmoveto{\pgfqpoint{0.000000in}{0.000000in}}%
\pgfpathlineto{\pgfqpoint{-0.048611in}{0.000000in}}%
\pgfusepath{stroke,fill}%
}%
\begin{pgfscope}%
\pgfsys@transformshift{0.750000in}{0.793323in}%
\pgfsys@useobject{currentmarker}{}%
\end{pgfscope}%
\end{pgfscope}%
\begin{pgfscope}%
\definecolor{textcolor}{rgb}{0.000000,0.000000,0.000000}%
\pgfsetstrokecolor{textcolor}%
\pgfsetfillcolor{textcolor}%
\pgftext[x=0.343556in,y=0.740562in,left,base]{\color{textcolor}\sffamily\fontsize{10.000000}{12.000000}\selectfont \(\displaystyle {-10^{1}}\)}%
\end{pgfscope}%
\begin{pgfscope}%
\pgfsetbuttcap%
\pgfsetroundjoin%
\definecolor{currentfill}{rgb}{0.000000,0.000000,0.000000}%
\pgfsetfillcolor{currentfill}%
\pgfsetlinewidth{0.803000pt}%
\definecolor{currentstroke}{rgb}{0.000000,0.000000,0.000000}%
\pgfsetstrokecolor{currentstroke}%
\pgfsetdash{}{0pt}%
\pgfsys@defobject{currentmarker}{\pgfqpoint{-0.048611in}{0.000000in}}{\pgfqpoint{0.000000in}{0.000000in}}{%
\pgfpathmoveto{\pgfqpoint{0.000000in}{0.000000in}}%
\pgfpathlineto{\pgfqpoint{-0.048611in}{0.000000in}}%
\pgfusepath{stroke,fill}%
}%
\begin{pgfscope}%
\pgfsys@transformshift{0.750000in}{1.460280in}%
\pgfsys@useobject{currentmarker}{}%
\end{pgfscope}%
\end{pgfscope}%
\begin{pgfscope}%
\definecolor{textcolor}{rgb}{0.000000,0.000000,0.000000}%
\pgfsetstrokecolor{textcolor}%
\pgfsetfillcolor{textcolor}%
\pgftext[x=0.343556in,y=1.407518in,left,base]{\color{textcolor}\sffamily\fontsize{10.000000}{12.000000}\selectfont \(\displaystyle {-10^{0}}\)}%
\end{pgfscope}%
\begin{pgfscope}%
\pgfsetbuttcap%
\pgfsetroundjoin%
\definecolor{currentfill}{rgb}{0.000000,0.000000,0.000000}%
\pgfsetfillcolor{currentfill}%
\pgfsetlinewidth{0.803000pt}%
\definecolor{currentstroke}{rgb}{0.000000,0.000000,0.000000}%
\pgfsetstrokecolor{currentstroke}%
\pgfsetdash{}{0pt}%
\pgfsys@defobject{currentmarker}{\pgfqpoint{-0.048611in}{0.000000in}}{\pgfqpoint{0.000000in}{0.000000in}}{%
\pgfpathmoveto{\pgfqpoint{0.000000in}{0.000000in}}%
\pgfpathlineto{\pgfqpoint{-0.048611in}{0.000000in}}%
\pgfusepath{stroke,fill}%
}%
\begin{pgfscope}%
\pgfsys@transformshift{0.750000in}{1.755635in}%
\pgfsys@useobject{currentmarker}{}%
\end{pgfscope}%
\end{pgfscope}%
\begin{pgfscope}%
\definecolor{textcolor}{rgb}{0.000000,0.000000,0.000000}%
\pgfsetstrokecolor{textcolor}%
\pgfsetfillcolor{textcolor}%
\pgftext[x=0.583333in,y=1.702874in,left,base]{\color{textcolor}\sffamily\fontsize{10.000000}{12.000000}\selectfont \(\displaystyle {0}\)}%
\end{pgfscope}%
\begin{pgfscope}%
\pgfsetbuttcap%
\pgfsetroundjoin%
\definecolor{currentfill}{rgb}{0.000000,0.000000,0.000000}%
\pgfsetfillcolor{currentfill}%
\pgfsetlinewidth{0.803000pt}%
\definecolor{currentstroke}{rgb}{0.000000,0.000000,0.000000}%
\pgfsetstrokecolor{currentstroke}%
\pgfsetdash{}{0pt}%
\pgfsys@defobject{currentmarker}{\pgfqpoint{-0.048611in}{0.000000in}}{\pgfqpoint{0.000000in}{0.000000in}}{%
\pgfpathmoveto{\pgfqpoint{0.000000in}{0.000000in}}%
\pgfpathlineto{\pgfqpoint{-0.048611in}{0.000000in}}%
\pgfusepath{stroke,fill}%
}%
\begin{pgfscope}%
\pgfsys@transformshift{0.750000in}{2.050991in}%
\pgfsys@useobject{currentmarker}{}%
\end{pgfscope}%
\end{pgfscope}%
\begin{pgfscope}%
\definecolor{textcolor}{rgb}{0.000000,0.000000,0.000000}%
\pgfsetstrokecolor{textcolor}%
\pgfsetfillcolor{textcolor}%
\pgftext[x=0.451581in,y=1.998230in,left,base]{\color{textcolor}\sffamily\fontsize{10.000000}{12.000000}\selectfont \(\displaystyle {10^{0}}\)}%
\end{pgfscope}%
\begin{pgfscope}%
\pgfsetbuttcap%
\pgfsetroundjoin%
\definecolor{currentfill}{rgb}{0.000000,0.000000,0.000000}%
\pgfsetfillcolor{currentfill}%
\pgfsetlinewidth{0.803000pt}%
\definecolor{currentstroke}{rgb}{0.000000,0.000000,0.000000}%
\pgfsetstrokecolor{currentstroke}%
\pgfsetdash{}{0pt}%
\pgfsys@defobject{currentmarker}{\pgfqpoint{-0.048611in}{0.000000in}}{\pgfqpoint{0.000000in}{0.000000in}}{%
\pgfpathmoveto{\pgfqpoint{0.000000in}{0.000000in}}%
\pgfpathlineto{\pgfqpoint{-0.048611in}{0.000000in}}%
\pgfusepath{stroke,fill}%
}%
\begin{pgfscope}%
\pgfsys@transformshift{0.750000in}{2.717948in}%
\pgfsys@useobject{currentmarker}{}%
\end{pgfscope}%
\end{pgfscope}%
\begin{pgfscope}%
\definecolor{textcolor}{rgb}{0.000000,0.000000,0.000000}%
\pgfsetstrokecolor{textcolor}%
\pgfsetfillcolor{textcolor}%
\pgftext[x=0.451581in,y=2.665186in,left,base]{\color{textcolor}\sffamily\fontsize{10.000000}{12.000000}\selectfont \(\displaystyle {10^{1}}\)}%
\end{pgfscope}%
\begin{pgfscope}%
\pgfsetbuttcap%
\pgfsetroundjoin%
\definecolor{currentfill}{rgb}{0.000000,0.000000,0.000000}%
\pgfsetfillcolor{currentfill}%
\pgfsetlinewidth{0.803000pt}%
\definecolor{currentstroke}{rgb}{0.000000,0.000000,0.000000}%
\pgfsetstrokecolor{currentstroke}%
\pgfsetdash{}{0pt}%
\pgfsys@defobject{currentmarker}{\pgfqpoint{-0.048611in}{0.000000in}}{\pgfqpoint{0.000000in}{0.000000in}}{%
\pgfpathmoveto{\pgfqpoint{0.000000in}{0.000000in}}%
\pgfpathlineto{\pgfqpoint{-0.048611in}{0.000000in}}%
\pgfusepath{stroke,fill}%
}%
\begin{pgfscope}%
\pgfsys@transformshift{0.750000in}{3.249588in}%
\pgfsys@useobject{currentmarker}{}%
\end{pgfscope}%
\end{pgfscope}%
\begin{pgfscope}%
\definecolor{textcolor}{rgb}{0.000000,0.000000,0.000000}%
\pgfsetstrokecolor{textcolor}%
\pgfsetfillcolor{textcolor}%
\pgftext[x=0.451581in,y=3.196826in,left,base]{\color{textcolor}\sffamily\fontsize{10.000000}{12.000000}\selectfont \(\displaystyle {10^{2}}\)}%
\end{pgfscope}%
\begin{pgfscope}%
\definecolor{textcolor}{rgb}{0.000000,0.000000,0.000000}%
\pgfsetstrokecolor{textcolor}%
\pgfsetfillcolor{textcolor}%
\pgftext[x=0.288001in,y=2.010000in,,bottom,rotate=90.000000]{\color{textcolor}\sffamily\fontsize{10.000000}{12.000000}\selectfont \(\displaystyle c_n\) (m\(\displaystyle ^{-2}\))}%
\end{pgfscope}%
\begin{pgfscope}%
\pgfpathrectangle{\pgfqpoint{0.750000in}{0.500000in}}{\pgfqpoint{4.650000in}{3.020000in}}%
\pgfusepath{clip}%
\pgfsetrectcap%
\pgfsetroundjoin%
\pgfsetlinewidth{1.003750pt}%
\definecolor{currentstroke}{rgb}{0.000000,0.000000,0.000000}%
\pgfsetstrokecolor{currentstroke}%
\pgfsetdash{}{0pt}%
\pgfpathmoveto{\pgfqpoint{0.961364in}{1.755635in}}%
\pgfpathlineto{\pgfqpoint{1.878602in}{1.755635in}}%
\pgfpathlineto{\pgfqpoint{2.676201in}{1.755635in}}%
\pgfpathlineto{\pgfqpoint{3.473799in}{1.755635in}}%
\pgfpathlineto{\pgfqpoint{4.271398in}{1.755635in}}%
\pgfpathlineto{\pgfqpoint{5.188636in}{1.755635in}}%
\pgfusepath{stroke}%
\end{pgfscope}%
\begin{pgfscope}%
\pgfsetrectcap%
\pgfsetmiterjoin%
\pgfsetlinewidth{0.803000pt}%
\definecolor{currentstroke}{rgb}{0.000000,0.000000,0.000000}%
\pgfsetstrokecolor{currentstroke}%
\pgfsetdash{}{0pt}%
\pgfpathmoveto{\pgfqpoint{0.750000in}{0.500000in}}%
\pgfpathlineto{\pgfqpoint{0.750000in}{3.520000in}}%
\pgfusepath{stroke}%
\end{pgfscope}%
\begin{pgfscope}%
\pgfsetrectcap%
\pgfsetmiterjoin%
\pgfsetlinewidth{0.803000pt}%
\definecolor{currentstroke}{rgb}{0.000000,0.000000,0.000000}%
\pgfsetstrokecolor{currentstroke}%
\pgfsetdash{}{0pt}%
\pgfpathmoveto{\pgfqpoint{5.400000in}{0.500000in}}%
\pgfpathlineto{\pgfqpoint{5.400000in}{3.520000in}}%
\pgfusepath{stroke}%
\end{pgfscope}%
\begin{pgfscope}%
\pgfsetrectcap%
\pgfsetmiterjoin%
\pgfsetlinewidth{0.803000pt}%
\definecolor{currentstroke}{rgb}{0.000000,0.000000,0.000000}%
\pgfsetstrokecolor{currentstroke}%
\pgfsetdash{}{0pt}%
\pgfpathmoveto{\pgfqpoint{0.750000in}{0.500000in}}%
\pgfpathlineto{\pgfqpoint{5.400000in}{0.500000in}}%
\pgfusepath{stroke}%
\end{pgfscope}%
\begin{pgfscope}%
\pgfsetrectcap%
\pgfsetmiterjoin%
\pgfsetlinewidth{0.803000pt}%
\definecolor{currentstroke}{rgb}{0.000000,0.000000,0.000000}%
\pgfsetstrokecolor{currentstroke}%
\pgfsetdash{}{0pt}%
\pgfpathmoveto{\pgfqpoint{0.750000in}{3.520000in}}%
\pgfpathlineto{\pgfqpoint{5.400000in}{3.520000in}}%
\pgfusepath{stroke}%
\end{pgfscope}%
\begin{pgfscope}%
\pgfsetbuttcap%
\pgfsetmiterjoin%
\definecolor{currentfill}{rgb}{1.000000,1.000000,1.000000}%
\pgfsetfillcolor{currentfill}%
\pgfsetfillopacity{0.800000}%
\pgfsetlinewidth{1.003750pt}%
\definecolor{currentstroke}{rgb}{0.800000,0.800000,0.800000}%
\pgfsetstrokecolor{currentstroke}%
\pgfsetstrokeopacity{0.800000}%
\pgfsetdash{}{0pt}%
\pgfpathmoveto{\pgfqpoint{4.234052in}{2.797317in}}%
\pgfpathlineto{\pgfqpoint{5.302778in}{2.797317in}}%
\pgfpathquadraticcurveto{\pgfqpoint{5.330556in}{2.797317in}}{\pgfqpoint{5.330556in}{2.825095in}}%
\pgfpathlineto{\pgfqpoint{5.330556in}{3.422778in}}%
\pgfpathquadraticcurveto{\pgfqpoint{5.330556in}{3.450556in}}{\pgfqpoint{5.302778in}{3.450556in}}%
\pgfpathlineto{\pgfqpoint{4.234052in}{3.450556in}}%
\pgfpathquadraticcurveto{\pgfqpoint{4.206274in}{3.450556in}}{\pgfqpoint{4.206274in}{3.422778in}}%
\pgfpathlineto{\pgfqpoint{4.206274in}{2.825095in}}%
\pgfpathquadraticcurveto{\pgfqpoint{4.206274in}{2.797317in}}{\pgfqpoint{4.234052in}{2.797317in}}%
\pgfpathclose%
\pgfusepath{stroke,fill}%
\end{pgfscope}%
\begin{pgfscope}%
\pgfsetbuttcap%
\pgfsetmiterjoin%
\definecolor{currentfill}{rgb}{1.000000,0.000000,0.000000}%
\pgfsetfillcolor{currentfill}%
\pgfsetlinewidth{0.000000pt}%
\definecolor{currentstroke}{rgb}{0.000000,0.000000,0.000000}%
\pgfsetstrokecolor{currentstroke}%
\pgfsetstrokeopacity{0.000000}%
\pgfsetdash{}{0pt}%
\pgfpathmoveto{\pgfqpoint{4.261830in}{3.289477in}}%
\pgfpathlineto{\pgfqpoint{4.539608in}{3.289477in}}%
\pgfpathlineto{\pgfqpoint{4.539608in}{3.386699in}}%
\pgfpathlineto{\pgfqpoint{4.261830in}{3.386699in}}%
\pgfpathclose%
\pgfusepath{fill}%
\end{pgfscope}%
\begin{pgfscope}%
\definecolor{textcolor}{rgb}{0.000000,0.000000,0.000000}%
\pgfsetstrokecolor{textcolor}%
\pgfsetfillcolor{textcolor}%
\pgftext[x=4.650719in,y=3.289477in,left,base]{\color{textcolor}\sffamily\fontsize{10.000000}{12.000000}\selectfont \(\displaystyle 2^{\mathrm{nd}}\)-order}%
\end{pgfscope}%
\begin{pgfscope}%
\pgfsetbuttcap%
\pgfsetmiterjoin%
\definecolor{currentfill}{rgb}{0.000000,0.000000,1.000000}%
\pgfsetfillcolor{currentfill}%
\pgfsetlinewidth{0.000000pt}%
\definecolor{currentstroke}{rgb}{0.000000,0.000000,0.000000}%
\pgfsetstrokecolor{currentstroke}%
\pgfsetstrokeopacity{0.000000}%
\pgfsetdash{}{0pt}%
\pgfpathmoveto{\pgfqpoint{4.261830in}{3.085620in}}%
\pgfpathlineto{\pgfqpoint{4.539608in}{3.085620in}}%
\pgfpathlineto{\pgfqpoint{4.539608in}{3.182842in}}%
\pgfpathlineto{\pgfqpoint{4.261830in}{3.182842in}}%
\pgfpathclose%
\pgfusepath{fill}%
\end{pgfscope}%
\begin{pgfscope}%
\definecolor{textcolor}{rgb}{0.000000,0.000000,0.000000}%
\pgfsetstrokecolor{textcolor}%
\pgfsetfillcolor{textcolor}%
\pgftext[x=4.650719in,y=3.085620in,left,base]{\color{textcolor}\sffamily\fontsize{10.000000}{12.000000}\selectfont \(\displaystyle 4^{\mathrm{th}}\)-order}%
\end{pgfscope}%
\begin{pgfscope}%
\pgfsetbuttcap%
\pgfsetmiterjoin%
\definecolor{currentfill}{rgb}{0.000000,0.500000,0.000000}%
\pgfsetfillcolor{currentfill}%
\pgfsetlinewidth{0.000000pt}%
\definecolor{currentstroke}{rgb}{0.000000,0.000000,0.000000}%
\pgfsetstrokecolor{currentstroke}%
\pgfsetstrokeopacity{0.000000}%
\pgfsetdash{}{0pt}%
\pgfpathmoveto{\pgfqpoint{4.261830in}{2.881763in}}%
\pgfpathlineto{\pgfqpoint{4.539608in}{2.881763in}}%
\pgfpathlineto{\pgfqpoint{4.539608in}{2.978985in}}%
\pgfpathlineto{\pgfqpoint{4.261830in}{2.978985in}}%
\pgfpathclose%
\pgfusepath{fill}%
\end{pgfscope}%
\begin{pgfscope}%
\definecolor{textcolor}{rgb}{0.000000,0.000000,0.000000}%
\pgfsetstrokecolor{textcolor}%
\pgfsetfillcolor{textcolor}%
\pgftext[x=4.650719in,y=2.881763in,left,base]{\color{textcolor}\sffamily\fontsize{10.000000}{12.000000}\selectfont \(\displaystyle 6^{\mathrm{th}}\)-order}%
\end{pgfscope}%
\end{pgfpicture}%
\makeatother%
\endgroup%

%% file: Conclusion.tex
\section{Conclusions and Future Work}
\label{chap:conclusions}
The purpose of this paper is to show that nonlocal properties of solid materials and the peridynamic kernel function should be determined by the microstructural properties if accurate wave dynamics are desired.  Because wave propagation plays such an important role in dynamic fracture, this is something that should be considered carefully by the peridynamics community when simulating pervasive fracture and fragmentation. Our model problem is a one-dimensional initial/boundary-value problem with periodic microstructure. We have laid out a multiscale analysis that results in a discrete peridynamic kernel that is used to define the macroscopic displacement. The resulting discrete peridynamic kernel turns out to be of a sign-changing type kernel which has also been utilized by Wildman \cite{wildmandiscrete} to reduce wave dispersion in linearized peridynamic models.  Wildman points out that the negative kernel values result in unstable solutions in fracture problems, something we did not consider here.  Numerical experiments were conducted using both the kernel we derive and a standard peridynamic kernel that has been used widely. The numerical results show that our peridynamic kernel achieves better accuracy and that increasing the order of the polynomial anzatz increases the accuracy further.  Additionally, our model is consistent with local homogenization theory in the limit of vanishing horizon.

Future work should include extension of the theory presented here to higher dimensions.  However, this could prove intractable, especially for nonperiodic microstructures.  Perhaps, machine learning techniques could be used to learn the most accurate  kernel function, with respect to desired quantities of interest, e.g. dispersion relations, fracture properties, etc., for a given microstructure.